\documentclass[sn-mathphys,Numbered]{sn-jnl}

\usepackage[T1]{fontenc}
\usepackage[utf8]{inputenc}
\usepackage[french]{babel}
\usepackage[autostyle]{csquotes}

\usepackage[version-1-compatibility]{siunitx}

\usepackage{physics}
\usepackage{graphicx}%
\usepackage{multirow}%
\usepackage{amsmath,amssymb,amsfonts}
\usepackage{amsthm}%
\usepackage{mathrsfs}%
\usepackage[title]{appendix}%
\usepackage{xcolor}%
\usepackage{textcomp}%
\usepackage{manyfoot}%
\usepackage{booktabs}%
\usepackage{algorithm}%
\usepackage{algorithmicx}%
\usepackage{algpseudocode}%
\usepackage{listings}%

\theoremstyle{thmstyleone}%
\usepackage{tensor}
\begin{document}

\title[Article Title]{Remarks on the general relativistic formulation of quantum theory\footnote{[English translation of: V. Bargmann, \textit{Bemerkungen zur allgemein-relativistischen Fassung der Quantentheorie}, {\em Sitzber. kgl.-preu{\ss}. Akad. Wiss. Berlin, Sitzung der phys.-math. Klasse} {\bf XXIV} (1932) 346--354. Translated by A. E. S. Hartmann, who is particularly grateful to Giovanna Colombo (Biblioteca di Scienza dell'Universit{\`a} degli Studi della Insubria, Como, Italy) for retrieving the original paper. (25 January, 2025)]}}


\author[1]{\fnm{V.} \sur{Bargmann}\footnote{Berlin, 1908 -- Princeton, 1989.}\; }



\affil[1]{
	\country{In Berlin} }

\maketitle
\vspace{-.5cm}
\begin{center}\small 
	(Submitted by Mr. Schr{\"o}dinger.)
\end{center}

\vspace{1.5cm}

Recently, E. Schr{\"o}dinger transcribed Dirac's theory of electrons into the system of the general theory of relativity\footnote{[Fn. 1, p.346] E. Schr{\"o}dinger, Diracsches Elektron im Schwerefeld I. These meeting reports, N. XI, p. 105, 1932. [Translator's note: Republication in \href{https://doi.org/10.1007/s10714-019-2626-y}{Gen. Rel. Grav. \textbf{52}, 4 (2020)}.]}. The following is a slightly different approach to this question, which does not change the factual content of the theory, but which differs in that point substitutions and similarity transformations are not coupled with each other, whereas the desired hermiticity of certain operators is ensured by the introduction of a special operator\footnote{[Fn. 2, p.346] See Eqs. (\ref{Eq10a})-(\ref{Eq10b}).}. 

\vspace{.5cm}
{\bf \textoneoldstyle.} Preliminary remarks on the special theory of relativity. Only real coordinates are used $(x^\text{\textzerooldstyle} = ct\,, x^\text{\textoneoldstyle} = x\,, x^\text{\texttwooldstyle} = y \,, x^\text{\textthreeoldstyle} = z  )$. $\mathring{g}_{ik}$ denotes the fundamental metric tensor of the special theory of relativity, i.e.
\begin{equation}\tag{\textoneoldstyle}\label{Eq1}
	\mathring{g}_{oo}= -\text{\textoneoldstyle},  \mathring{g}_{11}= \mathring{g}_{22}=\mathring{g}_{33}=+1; \quad \mathring{g}_{ik}= \text{\textzerooldstyle} \quad (i\neq k).
\end{equation}
We will choose \textfouroldstyle\, special \textfouroldstyle-row matrices $\mathring{\gamma}_k$ that we want to keep. They have to fulfil the equations\footnote{[Fn. 3, p.346] The cross $\dagger$ always denotes the transposed and complex conjugated matrix. The asterisk $\ast$ denotes the complex conjugated.}
\begin{equation}\tag{\texttwooldstyle}\label{Eq2}
	\mathring{\gamma}_i \mathring{\gamma}_k + \mathring{\gamma}_k \mathring{\gamma}_i = \text{\texttwooldstyle} \mathring{g}_{ik}\,,
\end{equation}

\begin{equation}\tag{\textthreeoldstyle}\label{Eq3}
	\mathring{\gamma}^\dagger_{\text{\textzerooldstyle}} = - \mathring{\gamma}_{\text{\textzerooldstyle}}\,;\quad 
		\mathring{\gamma}^\dagger_1 = \mathring{\gamma}_1\,,\quad 
			\mathring{\gamma}^\dagger_2 = \mathring{\gamma}_2\,,\quad 
				\mathring{\gamma}^\dagger_3 =  \mathring{\gamma}_3\,.
\end{equation}
If one defines the operators $\mathring{\gamma}^k$ in the usual way by $\mathring{g}^{kl} \mathring{\gamma}_l$, the Dirac equation for the free field case can be written in the form  
\begin{equation}\tag{\textfouroldstyle}\label{Eq4}
		\mathring{\gamma}^k \frac{\partial \psi}{\partial \mathring{x}^k} = \mu \psi. \qquad \left(\mu = \frac{2\pi m c}{h}\right)
\end{equation}
If $\psi$ is represented by the matrix $\left(\begin{smallmatrix}
	\psi_{\text{\textoneoldstyle}} \\ 
	\psi_{\text{\texttwooldstyle}} \\
	\psi_{\text{\textthreeoldstyle}} \\
	\psi_{\text{\textfouroldstyle}}
\end{smallmatrix}\right)$, and correspondingly $\psi^\dagger$ by 
$\left(\begin{smallmatrix}
	\psi^\ast_{\text{\textoneoldstyle}} \,, 
	\psi^\ast_{\text{\texttwooldstyle}} \,,
	\psi^\ast_{\text{\textthreeoldstyle}} \,,
	\psi^\ast_{\text{\textfouroldstyle}}
\end{smallmatrix}\right)$,
then the components of the current tensor are 
\begin{equation}\tag{\textfiveoldstyle}\label{Eq5}
	\mathring{S}^k = \psi^\dagger \mathring{\gamma}_{\text{\textzerooldstyle}} \mathring{\gamma}^k \psi = - \psi^\dagger \mathring{\gamma}^\dagger_{\text{\textzerooldstyle}} \psi = \varphi^\dagger\mathring{\gamma}^k \psi \,,
\end{equation}
when is set
\begin{equation}\tag{\textsixoldstyle}\label{Eq6}
	\varphi = \mathring{\alpha} \psi\,, \quad\mathring{\alpha} = - \mathring{\gamma}_{\text{\textzerooldstyle}}.
\end{equation}
Since all matrices $\mathring{\gamma}_{\text{\textzerooldstyle}}\mathring{\gamma}^k$ are Hermitian, the expressions for $ \mathring{S}^k$ are always real.\\

\vspace{.5cm}
{\bf 2.} New coordinates are introduced by an arbitrary linear substitution\footnote{[Fn. 1, p.347] It does not need to be a Lorentz transformation!} with constant coefficients,
\begin{equation}\tag{\textsevenoldstyle}\label{Eq7}
	x^k = \tensor{a}{^k_l} \mathring{x}^l\,; \quad \mathring{x}^k = \tensor{b}{_l^k} x^l 
\end{equation}
Equation (\ref{Eq4}) then becomes
\begin{equation*}
	\mu \psi = \mathring{\gamma}^k \tensor{a}{^l_k} \frac{\partial\psi}{\partial x^l} = \gamma^l \frac{\partial \psi}{\partial x^l} \,,
\end{equation*}
with unchanged $\psi$ and $\gamma^l = \tensor{a}{^l_k} \mathring{\gamma}^k$\,, $\gamma_l = \tensor{b}{_l^k} \mathring{\gamma}_k$\,, from which, because of (\ref{Eq2}), $\gamma_i \gamma_k + \gamma_k \gamma_i = \texttwooldstyle g_{ik}$ still follows. ($g_{ik}$ is the fundamental metric tensor in the new system.) For the current vector one obtains  
\begin{equation}\tag{\texteightoldstyle}\label{Eq8}
	S^k = \tensor{a}{^k_l} \mathring{S}^l = \tensor{a}{^k_l}\varphi^\dagger \mathring{\gamma}^l\psi = \varphi^\dagger\gamma^k \psi. 
\end{equation}
It holds\footnote{[Fn. 2, p.347] The quantity $\varphi$ introduced by P. A. M. Dirac ({\em Proc. Roy. Soc.} \href{https://doi.org/10.1098/rspa.1928.0056}{{\bf \textoneoldstyle\textoneoldstyle\texteightoldstyle}, \textthreeoldstyle\textfiveoldstyle\texttwooldstyle, \textoneoldstyle\textnineoldstyle\texttwooldstyle\texteightoldstyle}) does not satisfy Eq.(\ref{Eq6}). Dirac leaves the relationship between $\varphi$ and $\psi$ undetermined; however, if one wants to equate the expression he gives for the current vector with $S^k$ (Eq. \ref{Eq8}), one is led to the relationship $\varphi = -(\alpha \gamma^{\text{\textzerooldstyle}})' \psi^\ast$, resp. $\varphi^\ast = -(\alpha \gamma^{\text{\textzerooldstyle}})^\dagger \psi = -(\alpha \gamma^{\text{\textzerooldstyle}}) \psi$. (The accent is intended to indicate the transition to the transposed matrix.)} $\psi= \alpha \varphi$, with $\alpha = \mathring{\alpha}$; however, the connection between $\alpha$ and $\gamma_{\text{\textzerooldstyle}}$ (see Eq. (\ref{Eq6})) is, in general, lost. But since $\alpha \gamma^k = \tensor{a}{^k_l} \mathring{\alpha}\mathring{\gamma}^l$ \,, with real coefficients $\tensor{a}{^k_l}$, $\alpha \gamma^k$ remains Hermitian. $\alpha$ itself is skew-Hermitian.

Finally, if one applies a similarity transformation to $\gamma^k$, so that 
\begin{equation}\tag{\textnineoldstyle}\label{Eq9}
	\gamma'^k = S^{-1}\gamma^k S\,,
\end{equation} 
then one also sets
\begin{equation}\tag{\textnineoldstyle a}
	\psi' = S^{-1}\psi\,,
\end{equation}
\begin{equation}\tag{\textnineoldstyle b}
	\alpha' = S^\dagger \alpha S
\end{equation}
and
\begin{equation}\tag{\textnineoldstyle c}
	\varphi' = \alpha'\psi' = S^\dagger \varphi.
\end{equation}
Hence, the transformations (\ref{Eq9}) leaves the Dirac equation and the current vector (\ref{Eq8}) unchanged, as well as the characteristic properties of $\alpha$, namely
\begin{equation}\tag{\textoneoldstyle\textzerooldstyle a}\label{Eq10a}
	\alpha + \alpha^\dagger = \text{\textzerooldstyle}
\end{equation}
and
\begin{equation}\tag{\textoneoldstyle\textzerooldstyle b}\label{Eq10b}
	(\alpha\gamma^k)^\dagger = \alpha\gamma^k\footnote{[Fn. 3, p.347] V. Fock ({\em Z. Physik} \href{https://doi.org/10.1007/BF01339714}{{\bf \textfiveoldstyle\textsevenoldstyle}, \texttwooldstyle\textsixoldstyle\textoneoldstyle, \textoneoldstyle\textnineoldstyle\texttwooldstyle\textnineoldstyle}) denotes the Hermitian matrices $\alpha^\dagger\gamma^k$ with $\gamma^k$.} \quad \text{or} \quad \alpha\gamma^k + \tensor{\gamma}{^k^\dagger} \alpha = \text{\textzerooldstyle}.
\end{equation}
From (\ref{Eq10b}), it follows for the commutators $\tensor{s}{^\mu^\nu} = \frac{1}{2} ( \gamma^\mu \gamma^\nu - \gamma^\nu \gamma^\mu)$ that
\begin{equation}\tag{\textoneoldstyle\textoneoldstyle}\label{Eq11} 
	\alpha \tensor{s}{^\mu^\nu} +  \tensor{s}{^\mu^\nu^\dagger} \alpha = \text{\textzerooldstyle},
\end{equation}
and, because of the reality of $g_{kl}$, also that $\alpha \gamma_l + \gamma^\dagger \alpha =\text{\textzerooldstyle}$. 

\vspace{.5cm}
{\bf 3.} Transition to the general theory of relativity. Let the metric field $g_{ik}(x)$ be given. We set ourselves the task of finding operators $\gamma_k(x)$ such that
\begin{equation}\tag{\textoneoldstyle\texttwooldstyle }\label{Eq12}
	\gamma_i\gamma_k + \gamma_k\gamma_i = \text{\texttwooldstyle} g_{ik}\,.
\end{equation}
From each special $\gamma-$field satisfying the Eq. (\ref{Eq12}), the most general one is obtained by similarity transformations with arbitrary $S(x)$\footnote{[Fn. 1, p.348] E. Schr{\"o}dinger, {\em loc. cit.}, p.\textoneoldstyle\textzerooldstyle\textsevenoldstyle.}.

At each point, we choose a local, pseudo-orthogonal reference system whose quantities we label with index $\circ$. The following transformations may then apply to covariant vectors:
\begin{equation}\tag{\textoneoldstyle\textthreeoldstyle } 
	w_k = \tensor{b}{_k^l}(x) \mathring{w}_l\,; \quad \mathring{w}_k = \tensor{a}{^l_k}(x) w_l\,.
\end{equation}
In particular,
\begin{equation}\tag{\textoneoldstyle\textthreeoldstyle a}
	g_{ik} = \tensor{b}{_i^j} \tensor{b}{_k^l} \mathring{g}_{jl}\,.
\end{equation}
We now define a special solution of (\ref{Eq12}), which we will call the \enquote{distinguished $\gamma-$field\footnote{[Translator's note: the original reads 'das \enquote{ausgezeichnete $\gamma-$Feld}'.]}}, by the equations
\begin{equation}\tag{\textoneoldstyle\textfouroldstyle}\label{Eq14} 
	\gamma_l(x) = \tensor{b}{_l^k} (x) \mathring{\gamma}_k \,;\quad \mathring{\gamma}_k = \tensor{a}{^l_k} (x) \gamma_l (x) \,,
\end{equation}
and
\begin{equation}\tag{\textoneoldstyle\textfouroldstyle a}\label{Eq14a}
	\alpha = -\mathring{\gamma}_{\text{\textzerooldstyle}}\,.
\end{equation}
This satisfies both Eqs. (\ref{Eq10a})-(\ref{Eq10b}), and Eq. (\ref{Eq12}) as well.

\vspace{.5cm}
{\bf 4.} The differential equations for the distinguished $\gamma-$field result from Eq. (\ref{Eq14}). Namely, one finds
\begin{equation*}
	\frac{\partial \gamma_i}{\partial x^l} 
	= \frac{\partial \tensor{b}{_i^k}}{\partial x^l} \, \mathring{\gamma}_k
	= \frac{\partial \tensor{b}{_i^k}}{\partial x^l} \, \tensor{a}{^\mu_k} \gamma_\mu 
    = C^\mu_{il} \gamma_\mu \,.
\end{equation*}  
On the other hand, according to (\ref{Eq12}),
\begin{equation*}
	2\frac{\partial g_{ik}}{\partial x^l} 
	= \frac{\partial \gamma_i}{\partial x^l} \gamma_k 
	+ \gamma_i\frac{\partial \gamma_k}{\partial x^l} 
	+ \frac{\partial \gamma_k}{\partial x^l} \gamma_i
	+ \gamma_k\frac{\partial \gamma_k}{\partial x^l} 
	= C^\mu_{il} (\gamma_\mu \gamma_k + \gamma_k\gamma_\mu)
	+ C^\mu_{kl} (\gamma_i \gamma_\mu + \gamma_\mu\gamma_i),
\end{equation*}
consequently
\begin{equation*}
	2 ( \Gamma^\mu_{kl} g_{i\mu} + \Gamma^\mu_{il} g_{k\mu}) = 2 (C^\mu_{il} g_{k\mu} + C^\mu_{kl} g_{i\mu}  ).
\end{equation*}
With
\begin{equation*}
	\tensor{A}{^\mu_i_l} = C^\mu_{il} + \Gamma^\mu_{il},
\end{equation*}
is thus
\begin{equation*}
	\tensor{A}{^\mu_i_l} g_{k\mu} + \tensor{A}{^\mu_k_l} \tensor{A}{^\mu_i_\mu} =\text{\textzerooldstyle}  
\end{equation*}
or 
\begin{equation*}
	A_{kil} + A_{ikl} = \text{\textzerooldstyle},
\end{equation*}
\begin{equation}\tag{\textoneoldstyle\textfiveoldstyle}\label{Eq15}
	\frac{\partial \gamma_i}{\partial x^l} - \Gamma^\mu_{il} \gamma_\mu 
	= \tensor{A}{^\mu_i_l} \gamma_\mu = A_{\mu i l} \gamma^\mu\,. 
\end{equation}
The right-hand side of this differential equation can be written\footnote{\label{Fn.8}[Fn.1, p.349] E. Schr{\"o}dinger, \textit{loc. cit.}, Equations (\texteightoldstyle), (\textnineoldstyle), (\textoneoldstyle\textzerooldstyle), (\textoneoldstyle\textfiveoldstyle), (\textoneoldstyle\textsixoldstyle). } in the form
\begin{equation}\tag{\textoneoldstyle\textsixoldstyle}\label{Eq16} 
	A_{\mu il} \gamma^\mu = \Gamma_l \gamma_i - \gamma_i \Gamma_l\,,
\end{equation}
if you set
\begin{equation}\tag{\textoneoldstyle\textsevenoldstyle}\label{Eq17}
	\Gamma_l = \frac{1}{4} A_{\mu\nu l} s^{\mu\nu} + a_l \cdot I
\end{equation}
(for arbitrary $a_l$). At the same time, Eq. (\ref{Eq17}) is the most general solution of (\ref{Eq16}). Thus one finally gets (cf. Fn. \ref{Fn.8})
\begin{equation}\tag{\textoneoldstyle\texteightoldstyle}\label{Eq18}
	\frac{\partial \gamma_i}{\partial x^l } = \Gamma^{\mu}_{il} \gamma_\mu + \Gamma_l \gamma_i - \gamma_i \Gamma_l\,.
\end{equation}
If one moves to a new $\gamma-$field under a similarity transformation, say $\gamma'_i = S^{-1} \gamma_i S$, then Eq. (\ref{Eq18}) is retained if the $\Gamma$ is subject to the transformation
\begin{equation}\tag{\textoneoldstyle\textnineoldstyle}\label{Eq19}
	\Gamma'_l = S^{-1} \Gamma_l S - A_l\,,
\end{equation}
with
\begin{equation}\tag{\texttwooldstyle\textzerooldstyle}\label{Eq20}
	A_l = S^{-1} \frac{\partial S}{\partial x^l} = - \frac{\partial S^{-1}}{\partial x^l} S\,.
\end{equation}
In contrast, the relation (\ref{Eq15}) no longer holds in general.

Since $a_k = \frac{\text{\textoneoldstyle}}{\text{\textfouroldstyle}} \Tr \Gamma_k$, it follows from (\ref{Eq19}) and (\ref{Eq20}) that
\begin{equation}\tag{\texttwooldstyle\textoneoldstyle}\label{Eq21}
	a'_k = a_k - \frac{1}{4} \frac{\partial}{\partial x^k} (\log \det S) \,.
\end{equation}
The integrability conditions of (\ref{Eq18}) are\footnote{[Translartor's note: I have corrected a typo on the right-hand side of Eq.(\ref{Eq22}).]}
\begin{equation}\tag{\texttwooldstyle\texttwooldstyle}\label{Eq22}
\Phi_{kl}	\gamma_i - \gamma_i \Phi_{kl} = \tensor{R}{_k_l_i^\mu} \gamma_\mu \,,
\end{equation}
with
\begin{equation}\tag{\texttwooldstyle\textthreeoldstyle}\label{Eq23}
	\Phi_{kl} = \frac{\partial \Gamma_l}{\partial x^k} - \frac{\partial \Gamma_k}{\partial x^l} + \Gamma_l \Gamma_k - \Gamma_k \Gamma_l\,,
\end{equation}
\begin{equation}\tag{\texttwooldstyle\textfouroldstyle}\label{Eq24}
	\Phi_{kl} = -\frac{1}{4} \tensor{R}{_k_l_\mu_\nu} s^{\mu\nu} + f_{kl}\cdot I\,,
\end{equation}
and (cf. Fn. \ref{Fn.8})
\begin{equation}\tag{\texttwooldstyle\textfiveoldstyle}\label{Eq25}
	f_{kl} = \frac{\partial a_l}{\partial x^k} - \frac{\partial a_k}{\partial x^l}\,.
\end{equation}

Due to the reality of the coefficients $A_{\mu\nu l}$, one obtains from (\ref{Eq11}) and (\ref{Eq17}) [the following relation] for the distinguished $\gamma-$field,
\begin{equation*}
	\alpha \Gamma_k + \Gamma^\dagger_k \alpha = (a_k + a^\ast_k) \alpha.
\end{equation*}
Regarding the $a_k$, nothing can be inferred from the Eqs. (\ref{Eq16}) and (\ref{Eq17}). Because of their physical significance \textendash\, apart from one real factor, the $ia_k$ are the components of the electromagnetic vector potential\,\textendash\, it is, therefore, obvious to state
\begin{equation}\tag{\texttwooldstyle\textsixoldstyle}\label{Eq26}
		a_k + a^\ast_k = \text{\textzerooldstyle}\,.
\end{equation}
This gives us
\begin{equation}\tag{\texttwooldstyle\textsevenoldstyle}\label{Eq27}
	\alpha \Gamma_k + \Gamma^\dagger_k \alpha = \text{\textzerooldstyle}\,.
\end{equation}
(On the generalization for arbitrary $\gamma-$fields, see Eq. (\ref{Eq28}).)

The Equation (\ref{Eq26}) does not remain valid under similarity transformations; however, the additional [relation] (\ref{Eq21}) is irrelevant once it does not affect the field strengths (\ref{Eq25}). 

\vspace{.5cm}
{\bf 5.} Determination of $\alpha$ for a given $\gamma-$field. With $\alpha$, $c\cdot \alpha$ also satisfies Equations (\ref{Eq10a})-(\ref{Eq10b}) if $c$ is a real number. This shows that the real factor $c$ is also the only parameter still available when determing $\alpha$: one goes at every point of the local, pseudo-orthogonal system of reference and (by a suitable similarity transformation, if necessary) to the operators $\mathring{\gamma}^k$ (Equations (\ref{Eq2}) and (\ref{Eq3})). Then, one can easily see that there is only one solution given by $c\mathring{\gamma}_{\text{\textzerooldstyle}}$. A positive factor $c(x)$ is irrelevant, since one can get rid of it by a similarity transformation with $S(x) = \rho(x)\cdot$\textoneoldstyle\, $\left(\rho\cdot \rho^\ast = \frac{\text{\textoneoldstyle}}{c}\right)$, which leaves the $\gamma-$field unchanged.

Since we want to assume that $\alpha$ does not vanish anywhere\footnote{[Fn. 1, p.350] It then follows from the transformation properties of $\alpha$ and the Equation $\mathring{\gamma}^{\text{\texttwooldstyle}}_{\text{\textzerooldstyle}}=\text-{\textoneoldstyle}$ that $\alpha$ always has a reciprocal.}, $c$ must not become zero either. Because $c$ is real and it is assumed to be continuous, consequently it is always positive or always negative. The only significantly different solutions of Eqs. (\ref{Eq10a})-(\ref{Eq10b}) are, therefore, $\alpha$ and $-\alpha$. A decision between them is possible by the following determination, which takes into account the physical significance of $S^\text{\textzerooldstyle}$ as a probability density: $\alpha$ must be chosen so that the time component of the current vector is always positive, that is,
\begin{equation*}
	S^{\text{\textzerooldstyle}} = \varphi^\dagger \gamma^{\text{\textzerooldstyle}} \psi = -\psi^\dagger \alpha\gamma^{\text{\textzerooldstyle}} \psi > \text{\textzerooldstyle}\,. 
\end{equation*}
This means that the Hermitian operator $\alpha\gamma^{\text{\textzerooldstyle}}$ (according to the Eq. (\ref{Eq10b})) should be negative definite. (This leads to the Eq. (\ref{Eq6}) for the system $\gamma^{\text{\textzerooldstyle}}$.) If the chosen coordinate system is physically admissible in the sense that $g_{\text{\textzerooldstyle} \text{\textzerooldstyle}} <\text{\textzerooldstyle}$ and $\sum_{i,k=1,2,3}\, g_{ik} \,\xi^i\xi^k\,$ are positive definite, this determination can be carried on without contradiction. It is also invariant to a transformation [with respect] to another admissible coordinate system if only the time direction is retained, that is, if\footnote{[Fn. 1, p.351] We suppress the exact proof of this.} $\frac{\partial x'{}^{\text{\textzerooldstyle}}}{\partial x^\text{\textzerooldstyle}}>\text{\textzerooldstyle}$. Therefore, it can be said that the matrix $\alpha$ is essentially uniquely determined by a given $\gamma-$field.

The generalisation from Eq. (\ref{Eq27}) to arbitrary $\gamma-$fields is given by
\begin{equation}\tag{\texttwooldstyle\texteightoldstyle}\label{Eq28}  
	\alpha_{(k)} \equiv \frac{\partial \alpha}{\partial x^k} + \alpha \Gamma_k + \Gamma^\dagger_k \alpha = \text{\textzerooldstyle}.
\end{equation}
Under point substitutions (that is, ordinary coordinate transformations $x'{}^i = f^i(x^\text{\textzerooldstyle}, x^\text{\textoneoldstyle}, x^\text{\texttwooldstyle}, x^\text{\textthreeoldstyle})$), the matrices $\alpha_{(k)}$ behave like covariant vector components, since this is fixed\footnote{[Fn. 2, p.351] E. Schr{\"o}dinger, {\em loc. cit.}, sec. III.} for the $\Gamma_k$, by the similarity transformations of the $\alpha$ itself  according to Eqs. (\ref{Eq19}) and (\ref{Eq20}). Since the $\alpha_{(k)}$ vanish for the distinguished $\gamma-$field\,\textendash\, given that $\frac{\partial \alpha}{\partial x^k} = \text{\textzerooldstyle}$ (by Eq. (\ref{Eq14a})\footnote{[Translator's note: I have corrected a typo here.]}) and  $\alpha \Gamma_k + \Gamma^\dagger_k \alpha = \text{\textzerooldstyle}$ (Eq. (\ref{Eq27}))  \,\textendash\,, the Eq. (\ref{Eq28}) holds for any $\gamma-$field. 

Because\footnote{[Translator's note: I have introduced an asterisc on the numbering of Equation (\ref{Eq28*}) in order to distinguish it from (\ref{Eq28}).]} 
\begin{equation}\tag{\texttwooldstyle\texteightoldstyle$^\ast$}\label{Eq28*}
	\frac{\partial}{\partial x^k} \frac{\partial \alpha}{\partial x^l} 
	- 	\frac{\partial}{\partial x^l} \frac{\partial \alpha}{\partial x^k} 
	=  	\text{\textzerooldstyle}
\end{equation}
follows from (\ref{Eq28}),
\begin{equation}\tag{\texttwooldstyle\textnineoldstyle}\label{Eq29}
	\alpha \Phi_{kl} + \Phi^\dagger_{kl} \alpha = \text{\textzerooldstyle}.
\end{equation}
This relation (by taking (\ref{Eq11}) and (\ref{Eq24}) into account) is reduced to $(f_{kl} + f^\ast_{kl})\alpha = \text{\textzerooldstyle}$, thus
\begin{equation}\tag{\textthreeoldstyle\textzerooldstyle}\label{Eq30}
	f_{kl} + f^\ast_{kl}=\text{\textzerooldstyle}.
\end{equation}
Therefore, the field strengths are always real (see Eq. (\ref{Eq26})).

\vspace{.5cm}
{\bf 6.} Covariant quantities and its derivatives\footnote{[Fn. 3, p.351] See E. Schr{\"o}dinger, {\em loc. cit.}, \textsection\, \textfiveoldstyle.}. The matrix system $T^{\mu\nu\cdot\cdot}_{\rho\sigma\cdot\cdot}$ is called a  tensor operator of rank $(m+n)$ if $T$\footnote{\label{Fn.14}[Fn. 4, p.351] We often omit the indices as soon as this does not lead to misunderstandings.} behaves like a $m-$fold contravariant, $n-$fold covariant tensor under point substitutions, and changes to $S^{-\text{\textoneoldstyle}} T S$ under a similarity transformation of $\gamma$. Examples are $\gamma^k$, $s^{\mu\nu}$, $\Phi_{kl}$. 

Likewise, we also want to speak of a $\psi-$tensor or a $\varphi-$tensor (resp. an $\alpha-$tensor) if a large system of the character of $\psi$ or $\varphi$ (resp. $\alpha$), such as $\Lambda^{\mu\nu\cdot\cdot}_{\rho\sigma\cdot\cdot}$ (resp. $M^{\mu\nu\cdot\cdot}_{\rho\sigma\cdot\cdot}$), transforms as a tensor under point substitutions, and changes to $S^{-\text{\textoneoldstyle}} \Lambda$ or to $S^{\dagger} \Lambda$ (resp. $S^\dagger M S$) under a similarity transformation of $\gamma$.

For example, if $T$ is a tensor operator, so is $\Lambda=T\psi$ a $\psi-$tensor, $K=T^\dagger\varphi$ a $\varphi-$tensor, $M=\alpha T$ an $\alpha-$tensor. In general, $M$, as well as $M^\dagger$ and $M\pm M^\dagger$, always are an $\alpha-$tensor. Consequently, an equation of the form $M\pm M^\dagger=\text{\textzerooldstyle}$ is invariant with respect to $S-$transformations, and hence are Eqs. (\ref{Eq10a}), (\ref{Eq10b}), (\ref{Eq11}), and (\ref{Eq29}) as well. 

The further explanations are based on the following facts \,\textendash\, we give them without further details, because later we will only use the explicitly stated relations (\ref{Eq34}) to (\ref{Eq38}), which are easy to verify with the help of (\ref{Eq19}) and (\ref{Eq20})  \,\textendash\,: The variables $G$\footnote{[Fn. 1, p.352] See note \textfouroldstyle\, on page \textthreeoldstyle\textfiveoldstyle\textoneoldstyle. [Translator's note: Fn. \ref{Fn.14}.]} in question tranforms as tensors of rank $(m+n)$ under point substitutions and according to the scheme $G\rightarrow PGQ$ under $S-$transformations, where $P$ [denotes] the sequence $(\text{\textoneoldstyle}, S^{-\text{\textoneoldstyle}}, S^\dagger)$, and $Q$ the sequence $(\text{\textoneoldstyle}, S, (S^\dagger)^{-\text{\textoneoldstyle}})$. (This selection is determined by the transformation law of the $\Gamma_l$.) All combinations $(P,Q)$ occur. They are obtained by adding the reciprocal, or resp. the Hermitian adjoint quantities, to the quantities already given above.

The covariant derivatives to be defined  \,\textendash\, in order to be able to pronounce the product rule (\ref{Eq32}) simply, we always denote them, deviating from E. Schr{\" o}dinger's notation, as\footnote{\label{Fn.16}[Fn. 2, p.352] It should be explicitly pointed out that the operators $\nabla_k$ are not written down once and for all as certain polynomials in the $\frac{\partial}{\partial x^k}$, $\gamma^k$, etc., as is usual in quantum mechanics. Rather, just as in the absolute differential calculus of Riemannian geometry, the operator $\nabla_k$ is first determined by the quantity $G$ on which it acts. Thus, in the expression $\nabla_{\text{\textoneoldstyle}}(\nabla_{\text{\textoneoldstyle}} G)$, both operators are different from each other, because the operator on the right is a tensor quantity of rank $(m+n)$, whereas the one on the left is a tensor of rank $(m + (n+\text{\textoneoldstyle}))$. The operators used in the compositions $\nabla_k \psi$ and $\nabla_k\varphi$ also differ from each other (see Equations (\ref{Eq34a}) and (\ref{Eq36a})). } $\nabla_k G $ \,\textendash\, transform as tensors of rank $(m+ (n+ \text{\textoneoldstyle}))$, or likewise according to the scheme $\nabla_k G  \rightarrow P(\nabla_k G )Q$.

They are of the form\footnote{[Fn. 3, p.352] See note \textfouroldstyle\, on page \textthreeoldstyle\textfiveoldstyle\textoneoldstyle. [Translator's note: resp., Fn. \ref{Fn.14}.]}
\begin{equation}\tag{\textthreeoldstyle\textoneoldstyle}\label{Eq31}
	\nabla_k G = \mathring{\nabla}_k G + F(G, \Gamma_k)\,,
\end{equation}
where $F$ is linearly dependent on $G$, $\Gamma_k$, and $\Gamma^\dagger_k$, and $\mathring{\nabla}_k$ is the ordinary covariant derivative of Riemannian geometry\footnote{[Fn. 4, p.352] For example, $\mathring{\nabla}_k \tensor{v}{^\mu_\rho}= \frac{\partial \tensor{v}{^\mu_\rho}}{\partial x^k} + \Gamma^\mu_{\alpha k} \tensor{v}{^\alpha_\rho} - \Gamma^\alpha_{\rho k} \tensor{v}{^\mu_\alpha}$.}. The additive term $F$ is necessary here to ensure the transformation rule $\nabla_k G \rightarrow P(\nabla_k G) Q$. The following [property] always applies:
\begin{equation}\tag{\textthreeoldstyle\textoneoldstyle a}\label{Eq31a}
	\nabla_k (G^\dagger) = (\nabla_k G)^\dagger\,.
\end{equation}
For $c-$tensors ($P=Q=\text{\textoneoldstyle}$), $\nabla_k G = \mathring{\nabla}_k G$. Two quantities $G'$ and $G''$, which change to $P'G'Q'$ and $P'',G'',Q''$, respectively, under $S-$transformations, define an admissible quantity $G=G'G''$ with transformation law $G\rightarrow P'GQ''$, if $Q'P''=\text{\textoneoldstyle}$. Then, it always hold
\begin{equation}\tag{\textthreeoldstyle\texttwooldstyle}\label{Eq32}
	\nabla_k (G'G'') = (\nabla_k G')G'' + G' (\nabla_k G'')\,.
\end{equation}

The definitions now follow:
\begin{enumerate}
	\item[a)] For a tensor operator:
\begin{equation}\tag{\textthreeoldstyle\textthreeoldstyle}\label{Eq33}
	\nabla_k T = \mathring{\nabla}_k T + T \Gamma_k - \Gamma_k T\,.
\end{equation}
Accordingly, Eq. (\ref{Eq18}) means: $\nabla_l \gamma_i =\text{\textzerooldstyle}$. Likewise, one finds that $\nabla_l \gamma^i =\text{\textzerooldstyle}$.
	\item[b)] For a $\psi-$tensor, which, by the way, can be constructed from $\psi$ by any number of differentiatons:
\begin{equation}\tag{\textthreeoldstyle\textfouroldstyle}\label{Eq34}
	\nabla_k \Lambda = \mathring{\nabla}_k \Lambda - \Gamma_k \Lambda\,,
\end{equation}
and, in particular,
\begin{equation}\tag{\textthreeoldstyle\textfouroldstyle a}\label{Eq34a}
	\nabla_k \psi = \frac{\partial \psi}{\partial x^k} - \Gamma_k \psi \,.
\end{equation}
From (\ref{Eq34}) and (\ref{Eq34a}), it follows that\footnote{[Fn. 1, p.353] Written in detail, $\nabla_k (\nabla_l \psi) $ means $\left( \frac{\partial}{\partial x^l} - \Gamma_l\right)(\nabla_l \psi) - \Gamma^\alpha_{kl} (\nabla_\alpha \psi)$, thus Eq. (\ref{Eq34b}) says something different from Eq.(26) in E. Schr{\"o}dinger (see note \texttwooldstyle\,, p. \textthreeoldstyle\textfiveoldstyle\texttwooldstyle\, [Fn. \ref{Fn.16}, resp.].) }
\begin{equation}\tag{\textthreeoldstyle\textfouroldstyle b}\label{Eq34b}
	\nabla_k (\nabla_l \psi) - \nabla_l (\nabla_k \psi) = \Phi_{lk} \psi\,.  
\end{equation}
	\item[c)] For an $\alpha-$tensor:
\begin{equation}\tag{\textthreeoldstyle\textfiveoldstyle}\label{Eq35}
	\nabla_k M = \mathring{\nabla}_k M + M \Gamma_k + \Gamma^\dagger_k M\,,
\end{equation}
in particular, for $\alpha$ itself, 
\begin{equation}\tag{cf. \ref{Eq28}}
\nabla_k \alpha = \frac{\partial \alpha}{\partial x^k} + \alpha \Gamma_k + \Gamma^\dagger_k \alpha = \text{\textzerooldstyle}.
\end{equation}
From the product rule (\ref{Eq32}), therefore follows $\nabla_k (\alpha\gamma^l) =\text{\textzerooldstyle}$.\footnote{[Fn. 2, 353] Exactly the same Equation (24) by V. Fock, {\em loc. cit.}. Note, however, that the sign used there corresponds to our $\mathring{\nabla}_k$, and $\gamma^l$ corresponds to our $\alpha^\dagger \gamma^l$.}
	\item[d)] For a $\varphi-$tensor:
\begin{equation}\tag{\textthreeoldstyle\textsixoldstyle}\label{Eq36}
	\nabla_k N = \mathring{\nabla}_k N + \Gamma^\dagger_k N\,,
\end{equation}
thus
\begin{equation}\tag{\textthreeoldstyle\textsixoldstyle a}\label{Eq36a}
	\nabla_k\varphi = \nabla_k (\alpha\psi) = \alpha (\nabla_k \psi) = \frac{\partial \varphi}{\partial x^k} + \Gamma^\dagger_k \varphi \,,
\end{equation}
and
\begin{equation}\tag{\textthreeoldstyle\textsixoldstyle b}\label{36b}
	\nabla_k\varphi^\dagger = (\nabla_k \varphi)^\dagger = - (\nabla_k \psi^\dagger) \alpha = - (\nabla_k\psi)^\dagger \alpha = \frac{\partial \varphi^\dagger}{\partial x^k} + \varphi^\dagger \Gamma_k\,.
\end{equation}
If $\Lambda$ is a $\psi-$tensor, then $L=\varphi^\dagger\Lambda$ is a $c-$tensor. Because of (\ref{Eq32}), it holds
\begin{equation}\tag{\textthreeoldstyle\textsevenoldstyle}\label{Eq37}
	\nabla_k L = \mathring{\nabla}_k L = (\nabla_k \varphi^\dagger) \Lambda + \varphi^\dagger (\nabla_k \Lambda) \,.
\end{equation}
Finally, we mention the application of the rule (\ref{Eq32}) to the product of two tensor operators $A$ and $B$: 
\begin{equation}\tag{\textthreeoldstyle\texteightoldstyle}\label{Eq38}
	\nabla_k (A^{\mu\cdot\cdot}_{\rho\cdot\cdot} B^{\nu\cdot\cdot}_{\sigma\cdot\cdot}) 
	=
	\nabla_k ( A^{\mu\cdot\cdot}_{\rho\cdot\cdot}) B^{\nu\cdot\cdot}_{\sigma\cdot\cdot}
	+ A^{\mu\cdot\cdot}_{\rho\cdot\cdot} \nabla_k (B^{\nu\cdot\cdot}_{\sigma\cdot\cdot})\,.
\end{equation}

\end{enumerate} 

\vspace{.5cm}
{\bf 7.} The Dirac equation reads
\begin{equation}\tag{\textthreeoldstyle\textnineoldstyle}\label{Eq39}
	\gamma^k \nabla_k \psi = \mu \psi.
\end{equation}
It is invariant under both, point substitutions and $S-$transformations. A special case is the well-known \enquote{gauge invariance} of the Dirac equation. This is the transformation with the matrix $S=e^{-i\lambda}\cdot \text{\textoneoldstyle}$, in which $\gamma^k$ and $\alpha$ remain unchanged, while $\psi$ in $e^{-i\lambda} \psi$ and $\alpha_k$ in $\alpha_k + i \frac{\partial \lambda}{\partial x^k}$ are transformed.

Using (\ref{Eq10b}) and (\ref{36b}), a corresponding relationship from Dirac's equation for $\varphi$, namely 
\begin{align}
	\mu \varphi^\dagger &= -\mu\psi^\dagger\alpha = - (\nabla_k\psi^\dagger) \gamma^{k\dagger} \alpha = (\nabla_k\psi^\dagger) \alpha \gamma^k \nonumber\\
	\mu \varphi^\dagger &= - (\nabla_k \varphi^\dagger) \gamma^k.\tag{\textfouroldstyle\textzerooldstyle}\label{Eq40}
\end{align}
The current components are given by $S^k= \varphi^\dagger \gamma^k\psi$. Consequently, according to the product rule and given that $\nabla_l \gamma^k = \text{\textzerooldstyle}$, 
\begin{equation*}
	\nabla_l S^k = (\nabla_l \varphi^\dagger) \gamma^k \psi + \varphi^\dagger \gamma^k (\nabla_l \psi).
\end{equation*}
Therefore, because of (\ref{Eq39}) and (\ref{Eq40}), the divergence of the four-current must vanishes,
\begin{equation}\tag{\textfouroldstyle\textoneoldstyle}\label{Eq41}
	\nabla_k S^k = (\nabla_k \varphi^\dagger) \gamma^k \psi + \varphi^\dagger \gamma^k (\nabla_k \psi) = -\mu \varphi^\dagger \psi + \mu \varphi^\dagger \psi = \text{\textzerooldstyle}.
\end{equation}

It can be seen that the Dirac's equation can be obtained from the variational principle $\delta\int \sqrt{g} H\,dx_\text{\textzerooldstyle}\,dx_\text{\textoneoldstyle}\,dx_\text{\texttwooldstyle}\,dx_\text{\textthreeoldstyle} =\text{\textzerooldstyle}$, [with] $H= \varphi^\dagger\gamma^k (\nabla_k \psi) - \mu \varphi^\dagger \psi$, namely the variation of $\varphi$ gives Eq. (\ref{Eq39}), and the variation of $\psi$ \,\textendash\, it is better to start from $H^\ast = (\nabla_k \varphi^\dagger) \gamma^k \psi + \mu \varphi^\dagger\psi$ \,\textendash\, gives Eq. (\ref{Eq40}).

We break off here, because the further considerations would be completely in line with those of E. Schr{\"o}dinger.

\end{document}